\documentclass[multphys,vecphys]{svmult}

\usepackage{makeidx}     
\usepackage{graphicx}    
\usepackage{multicol}    

\def\etal{{\it et al.}\ }
\def\apjs{AAstrophys.J. Suppl.~}
\def\apj{Astrophys. J.~}
\def\apjl{Astrophys. J. Lett.~}

\def\aap{Astron. Astrophys.~}

\def\nat{Nature~}
\def\sci{Science~}
\def\aph{astro-ph~}

\def\EE#1{\times 10^{#1}}

\newcommand{\kms}{\ensuremath{\rm ~km~s^{-1}}}

\newcommand{\Msun}{\ensuremath{M_\odot}}
\def\Msunyr{~\rm M_\odot~\rm yr^{-1}}
\def\Mdot{\dot M}
\def\gsim{\!\!\!\phantom{\ge}\smash{\buildrel{}\over
    {\lower2.5dd\hbox{$\buildrel{\lower2dd\hbox{$\displaystyle>$}}\over
                                 \sim$}}}\,\,}

\begin{document}

\title*{Modeling the Radio and X-ray Emission of SN 1993J and
    SN 2002ap}
\titlerunning{The CSM of SN 1993J and SN 2002ap}
\author{Claes Fransson\inst{1} and Claes-Ingvar Bj{\"o}rnsson\inst{1}}
\institute{Department of Astronomy, Stockholm University, AlbaNova,
    SE-106 91 Stockholm, Sweden
\texttt{claes@astro.su.se, bjornsson@astro.su.se}}
%
%
\maketitle

\begin{abstract}
Modeling of radio and X-ray observations of supernovae interacting
with their circumstellar media are discussed, with special application
to SN 1993J and SN 2002ap. We emphasize the importance of including
all relevant physical mechanisms, especially for the modeling of the
radio light curves. The different conclusions for the absorption
mechanism (free-free or synchrotron self-absorption), as well as
departures from an $\rho \propto r^{-2}$ CSM, as inferred by some
authors, are discussed in detail. We conclude that the evidence for a
variation in the mass loss rate with time is very weak. The
results regarding the efficiencies of magnetic field generation and
relativistic particle acceleration are summarized.

\end{abstract}
\section{Introduction}
\label{CF-intro}
The interaction of supernovae (SNe) with their circumstellar medium
(CSM) offer important clues to both the nature of the SN progenitors,
the hydrodynamics of the explosion, the environment of the SN, and the
physics of high velocity shock waves. The now convincing connection of
Type Ic SNe and GRB's has also made the study of the SN environment
especially interesting. Basically, the standard picture of the SN
interaction with the surroundings is only a non-relativistic version
of the standard afterglow scenario (e.g., Chevalier, this Proceedings).  We
will discuss a few issues related to the SN interaction.  For a more
detailed exposition see the recent review in \cite{CF-CF03}.

\section{The standard model}
\label{CF-stand}

Supernova progenitors come in basically two flavors; extended red
supergiants, or compact, hot stripped stellar cores. The former are
thought to be related to Type IIP SNe, while the latter are most
likely related to Type Ib and Ic SNe. Type IIL, IIn and IIb SNe
probably represent a decreasingly massive hydrogen envelope. It is
also tempting to identify this as a sequence of increasing ZAMS mass,
with the Type IIP representing the most common low mass progenitors,
and the Type Ic's originating from the most massive Wolf-Rayet
stars. A complication 
is the fact that a
large fraction of stars are in close binary systems. Mass transfer
between the companions can in this case lead to strong mass loss even
for stars of comparatively low mass. 

Massive stars have in general strong stellar winds. On the main
sequence the blue supergiants have fast winds with 
a wind velocity of $500-3000$ \kms. As the
star evolves to the red supergiant stage the escape velocity decreases
dramatically and consequently the wind velocity to $10-30$ \kms. In
most of this stage the mass loss rate is $10^{-6}-10^{-5}
\Msunyr$. There are, however, indications, both from the observations
of SNe (see below), and from stellar evolution calculations
\cite{CF-Heg97} that the star in some cases can undergo a stage with a
superwind, similar to what occurs in AGB's, with mass loss rates of
$\sim 10^{-4} \Msunyr$. The duration of this stage must be very short,
$\sim 10^4$ yrs. Finally, if mass loss is important enough, the star
may evolve to the Wolf-Rayet (WR) stage, with mass loss rate $\sim
10^{-5} \Msunyr$ and a wind velocity of $1000-5000$ \kms, depending on
the evolutionary stage of the WR star \cite{CF-NL00}. For single stars
this occurs only for stars more massive than $\sim 22-40$ \Msun\ for
solar metallicity, depending on rotation \cite{CF-MM03}. In a binary
system this may, however, occur at considerably lower mass. Because
the CS density is $\rho_{\rm cs}=\dot M/(4\pi u_{\rm w} r^2)$, where
$\dot M$ is the mass loss rate, $u_{\rm w}$ the wind velocity and $r$
the distance from the star, the CS density into which the SN explodes
can differ by several orders of magnitude, depending on the
evolutionary stage in which it explodes.

The collision of the supernova ejecta with the surrounding gas
generates a strong shock wave, which expands with a velocity $20-30
\%$ larger than the maximum velocity of the ejecta. The temperature
behind this forward shock is $\sim 10^9$ K. The pressure behind the
shock will send a reverse shock back into the ejecta. Because of the
higher density, the velocity of the reverse shock will only be $V_{\rm
cs}/(n-2) \sim 500-1000$ \kms, depending on the density gradient of
the ejecta, $\rho_{\rm ej} \propto r^{-n}$ \cite{CF-C82a}. For polytropic
envelopes the outer parts of the ejecta have $n \sim 10$
\cite{CF-MM99}. Departures from such a structure may, however, lead to
both steeper and shallower gradients. For typical ejecta gradients the
temperature will only be $10^7-10^8$ K behind this shock.
Therefore, and because of the
high ejecta density, cooling will in general be important for the
reverse shock \cite{CF-CF94,CF-CF84}. This will create a thin shell of cool
gas between the reverse and CS shocks, often referred to as the cool,
dense shell. The forward shock will in contrast usually be adiabatic,
unless the CS density is very high.

During the last two decades radio, optical and X-ray observations have
greatly added to our understanding of the SN environment (see the
papers by Immler, Filippenko and Van Dyk). Of
these, the radio provides the cleanest signature of CS
interaction. Because of this, and also because the interpretation of
these observations have generated some confusion, we will discuss
these observations in some detail, as well as the complications going
into the analysis.

\subsection{Radio emission and absorption}

The radio emission arises as a result of relativistic electrons
accelerated in the shock region and emitting synchrotron radiation in
the magnetic field. The exact mechanism of the acceleration is not
well known, although a first order Fermi mechanism would be natural
across the shock. Particle acceleration may, however, also occur
behind the shock, close to the contact discontinuity, separating the
shocked ejecta and shocked CSM. The generation of the magnetic field
is not much better understood, although there has been recent progress
in this area. In particular, numerical simulations \cite{CF-FHHN03,CF-NHR03} based on ideas of Medvedev \& Loeb \cite{CF-ML99} for GRB's, have
shown that the Weibel instability, which is a particular version of
the two-stream instability, can give rise to a strong magnetic field
at the shock. Although the simulations have mainly been done in the
context of relativistic shocks, the same mechanism should work also
for non-relativistic shocks. An important issue for this type of
simulations is to determine the efficiency of conversion of
the thermal energy of the shock into magnetic field and relativistic
electron energy. The latter two are usually characterized by
$\epsilon_{\rm e}=u_{\rm e}/u_{\rm therm}$ and $\epsilon_{\rm
B}=u_{\rm B}/u_{\rm therm}$, where $u_{\rm therm}= 9/8 \rho
V_s^2$. Most likely, these parameters depend on both the shock
velocity, the CSM density and chemical composition.

The radio emission may be affected by 
free-free absorption from the surrounding CSM and
synchrotron-self absorption (SSA) by the same relativistic electrons that
emit the radiation. Assuming the relativistic electrons to be
injected behind the forward shock with an energy distribution given
by $dn_{\rm e}(\gamma)/d\gamma\propto\gamma^{-p}$ for
$\gamma_{\rm min}\leq\gamma\leq\gamma_{\rm max}$, the synchrotron
spectrum is given by $F_\nu = \pi R^2 S_\nu [1-\exp(-\tau_\nu)]$ where $S_\nu \propto \nu^{5/2}/B^{1/2}$ and $\tau_\nu \propto
B^{3/2+\alpha} N_{\rm e} \nu^{-\alpha-5/2}$.  Here $\alpha=(p-1)/2$ and
$N_{\rm e}$ is the column
density of relativistic electrons. In the case of negligible
cooling $N_{\rm e}\propto R n_{\rm e} \propto R u_{\rm e}$.  At
low frequencies the optically thick spectrum is given by 

\begin{equation}
F_\nu
\propto R^2 \nu^{5/2} B^{-1/2},
\end{equation}

\noindent and at high frequencies 

\begin{equation}
F_\nu \propto
R^2 N_{\rm e} B^{1+\alpha} \nu^{-\alpha}\propto R^3 n_{\rm e} B^{1+\alpha}
\nu^{-\alpha}.
\end{equation}

\noindent A fit to the spectrum, covering the peak frequency
($\nu\equiv\nu_{\rm peak}$),
therefore allows a determination of both the magnetic field and the
density of relativistic electrons, if the radius of the emitting
region is known. As we discuss below, this can usually be determined
from optical line widths or in rare cases from VLBI.

 From these expressions we can derive
an estimate of the brightness temperature at the peak of the
spectrum (i.e., where $\tau \sim 1$) given by 

\begin{equation}
T_{\rm b} \approx
8.2\times\EE{10}g(p)^{2/17}
(\epsilon_{\rm e}/\epsilon_{\rm B})^{2/17} (F_\nu /10^{29} {\rm erg~Hz}^{-1})^{1/17} {\rm K},
\end{equation}

\noindent independent
of the value of $\nu_{\rm peak}$. The value of $g(p)$ varies slowly with
the parameters specifying the energy distribution of the relativistic
electrons, for example, $g(2)=1/\ln (\gamma_{\rm max}/\gamma_{\rm min}$). The
corresponding Lorentz factor is 

\begin{equation}
\gamma_{\rm peak} \approx
1.0 \EE{2} (\epsilon_{\rm e}/\epsilon_{\rm B})^{2/17} (F_\nu 
/10^{29}~{\rm erg~Hz}^{-1})^{1/17},
\end{equation}

\noindent showing that $\gamma_{\rm peak}$ is fairly insensitive
to various parameters, and is expected to vary by less than an order of
magnitude.

The second possibility, free-free absorption, is decoupled from the
emission region, and only depends on the properties of the CSM. For an
$\rho \propto r^{-2}$ CSM density, the free-free optical depth is

\begin{equation}
\tau_{\rm ff}\propto \nu^{-2} (\Mdot/u_{\rm w})^2 T^{-3/2} R_{\rm s}^{-3}.
\end{equation}

\noindent Here, $R_{\rm s}=V_{\rm s} t$.
To estimate the relative importance of these mechanisms we determine
the frequency of optical depth unity. For free-free absorption, we obtain

\begin{equation}
\nu_{\rm ff} \propto (\Mdot/u_{\rm w}) T^{-3/4} V_{\rm s}^{-3/2} t^{-3/2}.
\end{equation}

\noindent With 

\begin{equation}
B^2/8\pi \propto \epsilon_{\rm B} \rho V_{\rm s}^2 \propto
\epsilon_{\rm B} (\Mdot/u_{\rm w})
V_{\rm s}^2/R_{\rm s}^2 \propto \epsilon_{\rm B} (\Mdot/u_{\rm w}) /t^2
\end{equation}

\noindent and $N_{\rm e} \propto \epsilon_{\rm e} R_{\rm s} B^2/\epsilon_{\rm B}$ we obtain for the 
SSA frequency

\begin{equation}
\nu_{\rm SSA} \propto(\epsilon_{\rm B} \epsilon_{\rm e})^{1/3}
(\Mdot/u_{\rm w})^{2/3} V_{\rm s}^{1/3} t^{-1}. 
\end{equation}

\noindent The ratio of these
is 

\begin{equation}
\nu_{\rm ff}/\nu_{\rm SSA} \propto  (\epsilon_{\rm B} \epsilon_{\rm
e})^{-1/3} \linebreak
(\Mdot/u_{\rm w})^{1/3} V_{\rm s}^{-11/3} T^{-3/4} t^{-1/2}.
\end{equation}

Therefore, the relative importance of the two absorption mechanisms
depends strongly on the wind velocity of the progenitor, the
efficiencies of producing relativistic electrons and the strength of
the magnetic field, the CSM temperature and especially the shock
velocity. The latter point has been emphasized by Chevalier \cite{CF-C98},
who find that most Type Ic SNe fall into the SSA category, because of
their high expansion velocities and high wind velocities.

For a consistent modeling it is crucial to include all
relevant energy loss mechanisms for the relativistic electrons. These
include the usual synchrotron losses, Compton losses on the
photospheric and/or the synchrotron photons, and Coulomb losses. While the
former two
mainly act on the high energy electrons, and steepen the spectrum, the
latter is most important at low energies, and lead to a flattening of
the spectrum. Even with a constant value for $p$, this will lead to a curved
optically thin spectrum.
In practice, the kinetic equation for
the electron distribution should be solved at each energy. More
details on these processes can be found in \cite{CF-FB98}.

When the cooling time starts to become comparable to the dynamical
time (i.e., $t$), radiation losses will affect the emitted synchrotron
spectrum. For synchrotron cooling

\begin{equation}
t_{synch}/t=0.47(t/10\,{\rm
days})(\gamma\,\epsilon_{\rm B})^{-1}(\dot
M/10^{-5}\,\Msunyr)^{-1} \linebreak (u_{\rm w}/10~{\rm km s}^{-1}).
\end{equation}

\noindent Since $\gamma\gsim 10^2$, synchrotron
cooling is expected to be important for supernovae with a red
supergiant progenitor, unless $\epsilon_{\rm B}\ll 1$. However, cooling in
supernovae with a Wolf-Rayet star progenitor cannot be excluded,
since the cooling time may be shortened due to inverse Compton losses.
This can occur in situations where $\epsilon_{\rm e}\gg\epsilon_{\rm B}$
(synchrotron self-Compton) or for very luminous supernovae for which
scattering of photospheric photons may become important. As mentioned
above, knowing $\nu_{\rm peak}$ and $R_{\rm s}$ makes it possible to
deduce individual values for $\epsilon_{\rm B}$ and $\epsilon_{\rm e}$. The
occurrence of cooling provides an extra constraint on the model; for
example, the underlying assumption of a spherically symmetric source
geometry can be tested. The source properties can be further
refined if the Compton scattered radiation is observed.
Furthermore, if a low frequency flattening due to Coulomb losses is 
not properly
account for, it is likely that model fitting will give a value of
$\nu_{\rm peak}$, and hence $B$, which is too large.

If free-free absorption is important, the epoch of optical depth unity
allows us to determine the important density parameter, $\Mdot/u_{\rm
w}$.  Because $R_{\rm s} = V_{\rm s} t$, one can determine the ratio
$\Mdot/u_{\rm w}$ from the time of optical depth unity at a given
frequency, {\it if} the shock velocity and temperature of
the CSM are known. The shock velocity can either be estimated from the
maximum extent of the optical line widths, usually the H$\alpha$
line, or in more rare cases, like for SN 1993J, directly from the
radius as inferred from VLBI observations. 

The temperature of the CSM is even more difficult to determine. At the
time of shock break-out a burst of soft X-rays heats and ionizes the
CSM. The spectrum and luminosity of this burst depends on the nature
of the progenitor, and has recently been discussed by Matzner \& McKee
\cite{CF-MM99}. A compact progenitor has a very brief burst of hard
radiation with a low total energy, while a red supergiant has a longer
burst, but considerably softer and with a larger energy. The ring of
SN 1987A offers a spectacular example of a circumstellar structure
heated and ionized by the outburst of the SN. In this case the initial
temperature of the ring was $\sim 2\times 10^5$ K, cooling on a time scale of
years \cite{CF-LF91,CF-LF96}. 
At the smaller distances relevant for the free-free
absorption of the radio supernovae during the first months
($10^{15}-10^{16}$ cm), these effects will be even larger.

Unfortunately, with the exception of SN 1987A, little direct
information on the temperature of the CSM is available. Instead, one
has to rely on theoretical calculations. Because of the low density
the recombination and ionization time scales are long compared to the
evolution of the ionizing spectrum and one has to calculate the
evolution of the temperature and ionization of the CSM from the
out-break until the epoch of interest. This has only been done for a
few cases, the Type IIL SN 1979C \cite{CF-LF88}, SN 1987A
\cite{CF-LF91,CF-LF96} and SN 1993J \cite{CF-FLC96}. In all cases the CSM
temperatures close to the shock are $10^5-10^6$ K immediately after
shock break-out. The temperature then decreases on a time scale of
months, but can be $\gsim 2\EE4$ K even a year after the
explosion. This result is in sharp contrast to the simplified analyzes
of the radio light curves which assume a {\it constant} temperature of
in most cases $\sim
(1-3)\times 10^4$ K, and obviously leads to large errors in the estimated mass
loss rates. Moreover, the decreasing temperature can mimic a change in
the mass loss rate with time, as has been claimed for some of the
radio supernovae. These effects are discussed in detail in
\cite{CF-LF88}.

\section{SN 1993J}
\label{CF-sn93j}

\subsection{Synchrotron self-absorption or free-free absorption?}

The radio observations of SN 1993J \cite{CF-CRB04,CF-PT02,CF-vD94} are to date by far
the best ones available. Equally important, the VLBI observations
\cite{CF-B00,CF-M97} give direct information about the size of the radio
emitting region, and therefore the shock velocity, as function of
time. From this combination a detailed modeling of the spectra have
been possible \cite{CF-FB98,CF-FLC96,CF-PT01,CF-vD94}. The conclusions
with regard to the mechanism behind the absorption, as well as the
structure of the CSM, however, differ substantially.

The analysis in both \cite{CF-FLC96} and \cite{CF-vD94} were based on
modeling the radio observations by free-free absorption in a CSM. As
was shown in \cite{CF-FB98}, this alone cannot reproduce the turn-off of
the spectrum at low frequencies, which, especially at late epochs, is
considerably shallower than the exponential cutoff expected for
free-free absorption. Instead, it agrees well with the standard $F_\nu
\propto \nu^{5/2}$ spectrum expected from synchrotron
self-absorption. Only at early epochs was free-free absorption found
to be important, and then only for the longest wavelengths.

The main reason why free-free is only marginally important is that the
temperature of the CSM inferred from modeling of the heating and
ionization by the radiation from the shock break-out is likely to have
been very high, $\sim 10^5-10^6$ K \cite{CF-FLC96}. The parametric fits
by Van Dyk et al., on the other hand do not determine the temperature
either observationally or theoretically, and simply assume a constant
temperature. Because $\tau_{\rm ff}\propto T^{-3/2}$, this partly
explains the different importance of the free-free absorption in
\cite{CF-vD94} and \cite{CF-FB98}. Furthermore, Van Dyk et al. do not
use the dynamical evolution from the VLBI observations. Finally,
their fits do not include any cooling processes which affect the
electron spectrum, and only assume a constant power law for this. Both
Coulomb and synchrotron were shown to be
important in \cite{CF-FB98}, which flatten the low energy spectrum and
steepen it at high energies. The effect of this can be seen especially in
the lack of agreement between the fits and the observations
at the peak of the lowest frequency, caused by Coulomb
cooling, which flattens the electron spectrum.

The modeling in \cite{CF-PT01} is also based on SSA. They do, however, not
include a self-consistent calculation of the electron spectrum, for
example Coulomb cooling is not accounted for, and
invoke an arbitrary cutoff of the electron spectrum at low energies. This
explains the different quantitative conclusions from \cite{CF-FB98}; in
particular, the neglect of Coulomb cooling in the model fitting
causes the deduced value of the
synchrotron self-absorption frequency to be too high. This, in turn,
results in an overestimation of the strength of the magnetic field.

A test of the SSA model was provided by the new low frequency
observations with VLA in \cite{CF-PT02} and the VLA and GMRT in
\cite{CF-CRB04}. The fluxes of both these sets of observations agreed
well with the fluxes predicted in \cite{CF-FB98}. In addition, the
combined VLA and GMRT spectrum at 3200 days in \cite{CF-CRB04} nicely
showed the break in the spectrum caused by the synchrotron cooling to
evolve as expected.

One of the most interesting results of the modeling of SN 1993J was
that not only could individual values for the energy densities in
magnetic field and relativistic particles be derived but also their
evolution with time could be determined (see Fig. 1). Both of these
energy densities scaled with the thermal energy density behind the
shock (i.e., $\epsilon_{\rm B}$ and $\epsilon_{\rm e}$ are constants)
but the conditions are far from equipartition since $\epsilon_{\rm
B}\approx 0.14$ and $\epsilon_{\rm e}\sim 5\times10^{-4}$.
\begin{figure}[t]
\begin{center}
\includegraphics[scale=0.3,origin=c]{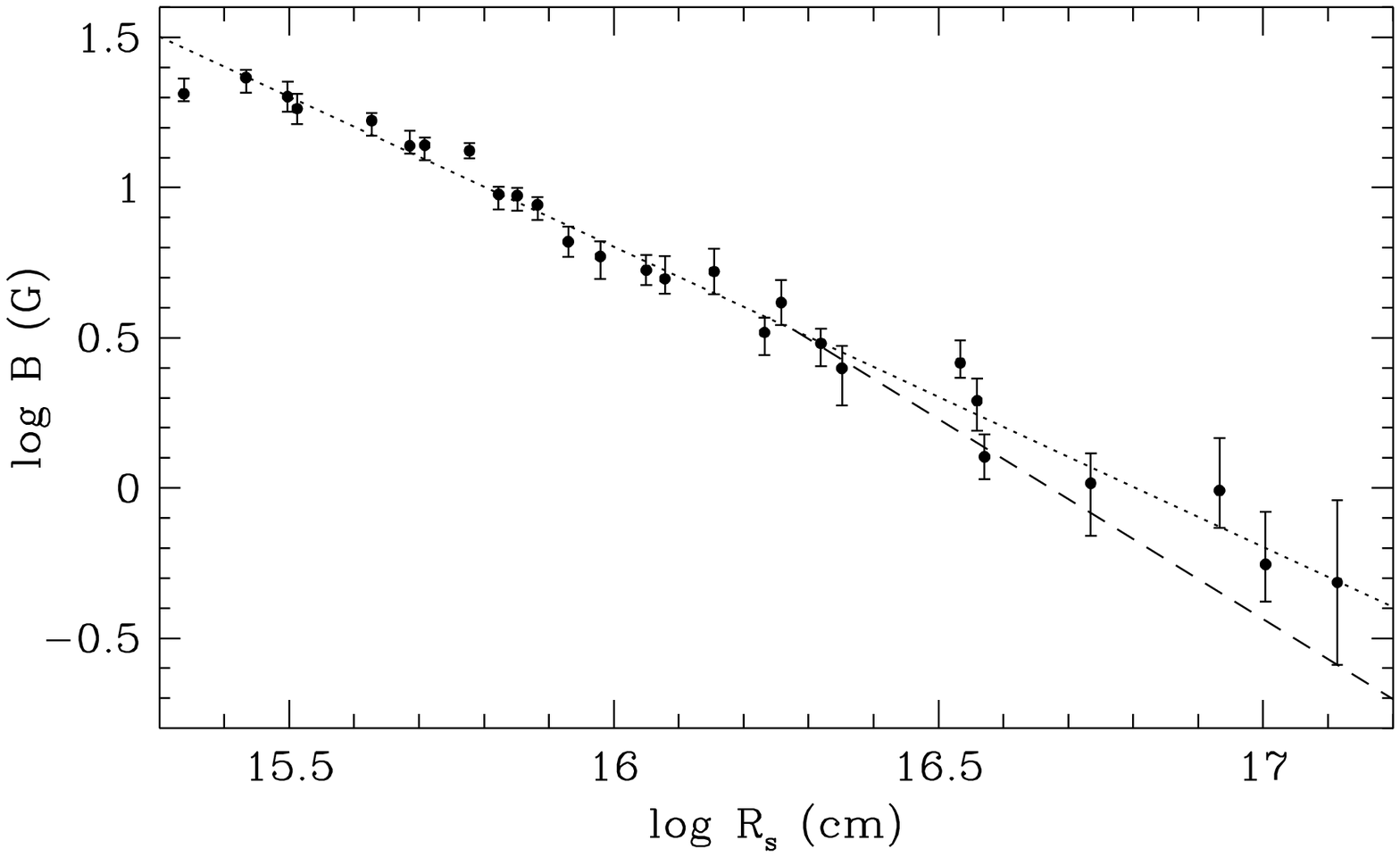}
\includegraphics[scale=0.3,origin=c]{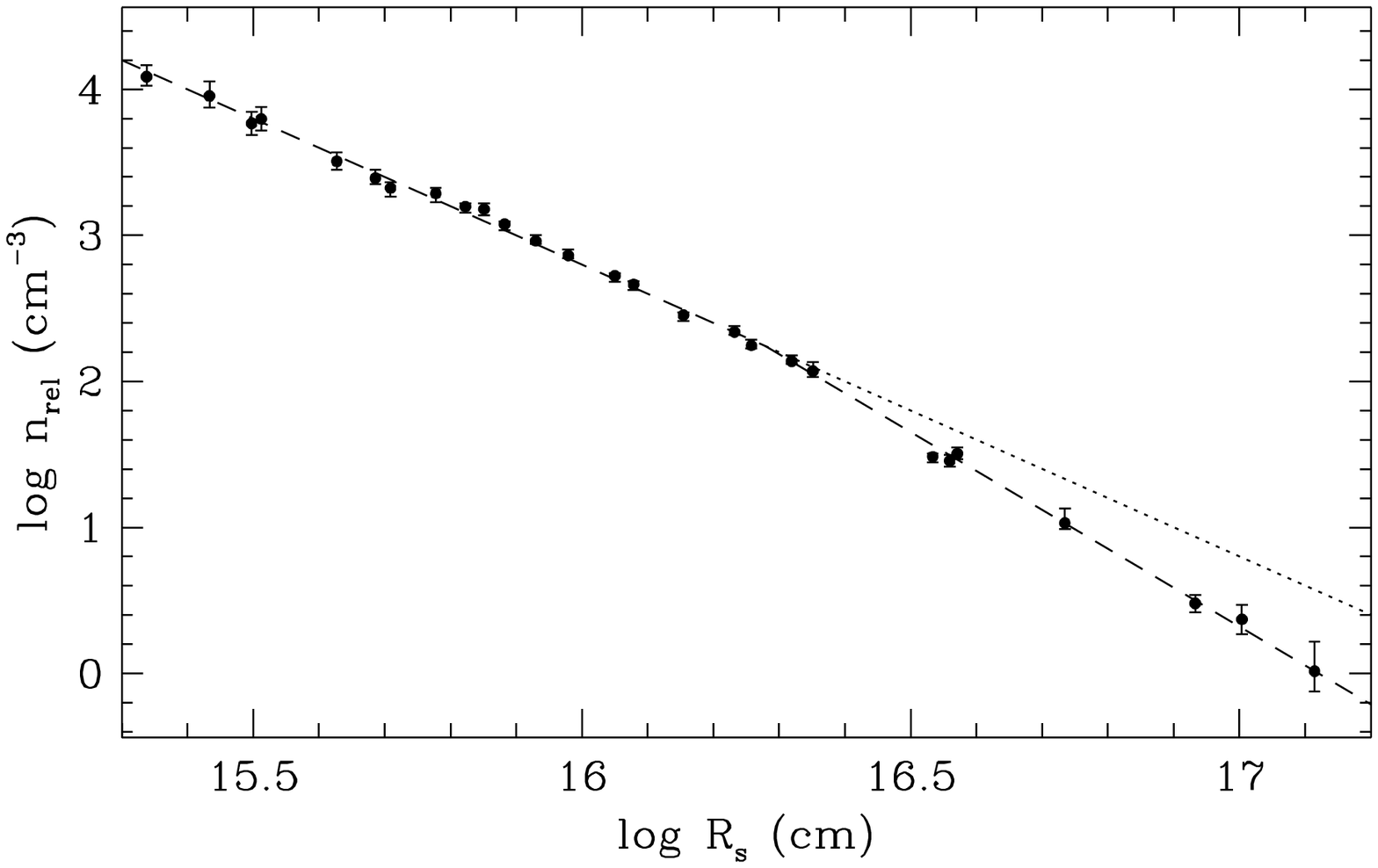}
\end{center}
\caption[]{Magnetic field (left) and density of relativistic electrons
(right) as a function of the shock radius for SN 1993J. The dashed
lines show the expected evolution if the magnetic energy density and
relativistic particle density scale with the thermal energy density,
$B^2/8 \pi \propto \rho V_{\rm s}^2 \propto n_{\rm rel} \propto
t^{-2}$, while the dotted lines show the case when $B \propto r^{-1}$
and $n_{\rm rel} \propto r^{-2}$\cite{CF-FB98}.}
\label{CF-fig4}
\end{figure}

\subsection{X-rays}

X-ray emission from SN 1993J were observed with OSSE on Compton/GRO
\cite{CF-Leis94}, ROSAT \cite{CF-IAW01,CF-Zim94}, ASCA \cite{CF-Uno02}, Chandra
\cite{CF-Swar03} and XMM \cite{CF-Zim03}. During the first two months,
the OSSE and ASCA observations showed a very hard spectrum with $kT
\sim 100$ keV. This agreed well with that expected from the forward,
circumstellar shock $\sim 1.1\EE2 (V_{\rm s}/10^4$ \kms)$^2$ keV \cite{CF-FLC96}.

Because of the proximity to the Sun, it could then not be observed
until $\sim 200$ days after explosion. 
The temperature was now
only $\sim 1$ keV, and the column density had increased by a large
factor \cite{CF-Uno02,CF-Zim94}. This transition was explained in
\cite{CF-FLC96} as a natural consequence of the radiative reverse shock
and the presence of the cool, dense shell. As the SN expanded the
column density of the CDS decreases $\propto t^{-1}$, and therefore
gradually becomes optically thin. Because the luminosity below $\sim
10$ keV is dominated by the reverse shock, there will be a hard to
soft transition, once the CDS becomes transparent.
 From the X-ray flux the mass loss rate was estimated to be
$\sim4\EE{-5} \Msunyr$, in agreement with that estimated from the
radio.

\subsection{Structure of the CSM}
There has been considerable confusion also with regard to the
structure of the CSM of SN 1993J. In several papers
\cite{CF-FLC96,CF-IAW01,CF-SN95,CF-vD94} there have been claims of a CSM density
varying as $\rho \propto r^{-1.5-1.7}$. This has in turn been taken as
evidence for a mass loss varying with time. Only in \cite{CF-FB98} was an
$\rho \propto r^{-2}$ found to reproduce the observations. It is
therefore of interest to examine the arguments on which these quite
different conclusions rest.

As has already been discussed, the analysis in \cite{CF-FLC96} and
\cite{CF-vD94} both neglected the effects of SSA, and are therefore not
physically consistent. Their conclusions from the radio modeling with
regard to the CSM should therefore be ignored.

The X-ray analysis by Immler et al. \cite{CF-IAW01} assumes that the X-ray
emission emerges from the forward, circumstellar shock. This is
directly contradicted by the low temperature, $\sim 1$ keV, and high
column density found from the ASCA observations later than
200 days \cite{CF-Uno02}. 
As was discussed above, the X-ray flux and
temperature at late phases are instead consistent with those expected from the
reverse shock \cite{CF-FLC96,CF-SN95}. 

The analysis by Suzuki \& Nomoto \cite{CF-SN95} is based on a
hydrodynamical, consistent modeling of the interaction of the ejecta
and CSM, and the results should therefore be taken seriously. The fact
that they obtain a CSM density at small radii varying as $\rho \propto
r^{-1.7}$ and a clumpy medium at large radii, however, depends on the
ejecta structure they use. The specific model they use, 4H47, was
designed to reproduce the early light curve, and had for this purpose
to be mixed artificially. More detailed modeling \cite{CF-Iwa97} also
showed that these 1-D models are hydrodynamically
unstable. A further problem with this model is that it does not
reproduce the velocity evolution of the shock, as inferred by the VLBI
observations. More detailed calculations of the X-ray evolution using
an ejecta model which does reproduce the VLBI observations indeed show
that a satisfactory reproduction of the X-ray observations can be
obtained for an $\rho \propto r^{-2}$ CSM density.

In conclusion, {\it the only self-consistent modeling of the CSM of SN
1993J are those in \cite{CF-FB98} using radio observations and
\cite{CF-SN95} using X-rays}. Because of the insensitivity of the radio
observations to the ejecta structure, in contrast to the X-rays, we
believe the former is the more reliable, and that an $\rho \propto
r^{-2}$ CSM density is the most likely.

\section{SN 2002ap}
\label{CF-sn2002ap}

Type Ic SNe are of special interest because of their relation to the
GRBs. 
The recent SN 2002ap is
together with SN 1998bw the best observed objects of this class, and we
discuss some issues related to this.

A special property of the Type Ic SNe is their high expansion
velocities. Although this varies by a large factor from the relatively
slow SN 1994I to SN 1998bw, they are in all cases much faster than the
Type II's and also SN 1993J.  A problem here is the fact that the
expansion velocity of the radio emitting region is difficult to
estimate directly from observations. The optical spectrum shows few
clear line features at early time, implying that blending is very
important. This only allows a lower limit to the velocity to be
determined. For SN 1998bw this was $\sim 60,000$ \kms \cite{CF-Nak01}, and
for SN 2002ap $\sim 30,000$ \kms \cite{CF-Maz02}.

Because of the high expansion velocities SSA dominates the radio
absorption. The radio observations of SN 1998bw \cite{CF-Kul98} have been
discussed in \cite{CF-LC99}. The most important conclusions was the high
expansion velocity, $V_{\rm s}/c \gsim 0.9$, and the large energy in
relativistic ejecta. Unfortunately, the radio observations of this and
other Type Ic's did not allow a determination of the mass loss rate,
because this depends on the unknown values of $\epsilon_{\rm e}$ and
$\epsilon_{\rm B}$. While these are often assumed to be close to
equipartition, there is little theoretical motivation for this.

SN 2002ap is interesting in this respect because, in addition to good
radio observations \cite{CF-BKC02}, it was also observed with XMM
\cite{CF-SPM03,CF-SCB03}. The X-ray emission is most naturally explained as
a result of inverse Compton scattering of the photospheric radiation by
the same relativistic electrons which are responsible for the radio
emission \cite{CF-BF04}. This process also leads to cooling of the
relativistic electrons, and thereby a steepening of the emitted
spectrum, which agrees with the fact that the optically thin radio
spectrum is as steep as $F_\nu \propto \nu^{-0.9}$.

By modeling the radio light curves of the different frequencies one
can determine the expansion velocity of the radio emitting region, as
well as the relative values of $\epsilon_{\rm e}$ and
$\epsilon_{\rm B}$. We find that $V_{\rm s} \sim 70,000$ \kms, and that
$\epsilon_{\rm e}$ and $\epsilon_{\rm B}$ are roughly equal (i.e., the
magnetic field and relativistic electrons are in approximate
equipartition). The exact value of their ratio depends on the
upper and lower
cut-offs in the electron distribution, which are not directly observable.
This is in contrast to SN 1993J, where we found that
$\epsilon_{\rm B}\gg\epsilon_{\rm e}$. It is also interesting to note
that in both 1993J and 2002ap the injected distribution of
relativistic electrons has $p\approx 2$, which is the theoretical
value expected from first order Fermi-acceleration at a
non-relativistic shock. This is not directly apparent from the
observations, since cooling alters the emitted spectrum so that the
simple relation $\alpha=(p-1)/2$ cannot be used to deduce the value
of $p$. It could be that an electron distribution with $p=2$ is valid
for most supernovae and that the range of observed values for $\alpha$
is due to cooling.

\section{Conclusions}
\label{CF-conc}

The VLA radio observations of SN 1993J are unique in terms of both the
temporal coverage and the quality of the observations. The fact that
also VLBI observations exist adds to this characterization, and means
that the size of the radio emitting plasma can be determined without
further assumptions. This combination makes a detailed spectral
analysis possible, and from this a determination of the magnetic field
and relativistic electron density as functions of time. For this a
consistent physical model is necessary, which includes both the
radiative transfer and the effects of different energy loss processes
on the electron spectrum. There is no need to invoke arbitrary
parameterized models, which only hides the real physical parameters,
and may result in misleading conclusions with regard to physical
mechanisms as well as the structure of the CSM.

\acknowledgement

We are grateful to Peter Lundqvist and Roger
Chevalier for comments and collaborations on several of these issues.



\end{document}